\documentclass[namedreferences]{kluwer}
\usepackage{graphicx}
\begin{document}                                                                                   
\begin{article}
%
%
\begin{opening}         
\title{Comparing peanut--shaped `bulges' to
       N--body\\ simulations and orbital calculations}
\runningtitle{Peanut--shaped `bulges' in edge--on galaxies}
\author{G.\ \surname{Aronica}$^1$, E.\ \surname{Athanassoula}$^2$,
M.\ \surname{Bureau}$^3$, A.\ \surname{Bosma}$^2$,
R.--J.\ \surname{Dettmar}$^1$, D.\ \surname{Vergani}$^4$,
M.\ \surname{Pohlen}$^5$}  
\runningauthor{Aronica et al.}
\institute{
$^1$ Astronomisches Institut, Ruhr--Universit\"at Bochum, Germany\\
$^2$ Observatoire de Marseille, France\\
$^3$ Hubble Fellow, Columbia Astrophysics Laboratory, Columbia University, USA\\
$^4$ Radio Astronomisches Institut, Universit\"at Bonn, Germany\\
$^5$ Instituto de Astrof\'{\i}sica de Canarias, Spain
}
\begin{abstract} 
We present a near-infrared $K_n$--band photometric study of edge--on
galaxies with a box/peanut--shaped `bulge'. The morphology of the
galaxies is analysed using unsharp masking and fits to the vertical
surface brightness profiles, and the results are compared to N--body
simulations and orbital calculations of barred galaxies. Both
theoretical approaches reproduce the main structures observed.
\end{abstract}
\end{opening}           
\section{Introduction} 
\begin{figure}
\centerline{\includegraphics[width=23pc,angle=0]{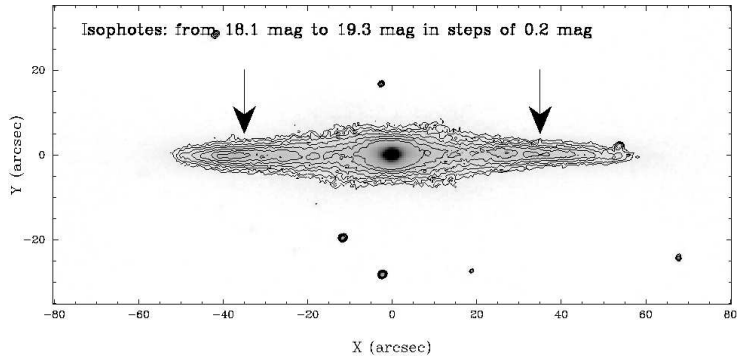}}
\caption[]{Image of ESO~443~-042, a galaxy with a peanut--shaped `bulge'
in the inner $15$~arcsec. Isophotes are overlaid on a grayscale
rendition of the $K_n$-band image. The arrows point to local maxima of
the surface brightness along the major axis.}
\label{fig:contour}
\end{figure}

Box/peanut (b/p) shaped `bulges' are frequently observed in edge--on
galaxies (see Fig.~\ref{fig:contour}; \citeauthor{ldp00a}
\citeyear{ldp00a} and references therein). They are also seen in many
N--body simulations of barred disks which display a vertical
thickening of the bar during its evolution, with a projected shape
similar to the observed b/p structures (e.g.\ \citeauthor{cdfp90}
\citeyear{cdfp90}; Athanassoula and Misiriotis \citeyear{am02},
hereafter \citeauthor{am02}). The works of \inlinecite{mk99} and
Bureau and Freeman (\citeyear{bf99}, hereafter \citeauthor{bf99})
support this view observationally. They analysed ionised gas
position--velocity diagrams of edge--on spirals using the method
described in \inlinecite{km95} and \inlinecite{ab99}. Kinematic bar
signatures were detected in essentially all the galaxies showing a b/p
feature. In a large sample of galaxies ($N=60$), the photometric link
between bars and b/p `bulges' was demonstrated reliably for the first
time by \citeauthor{ldp00b} (\citeyear{ldp00b}).
\section{Data} 
We analysed near-infrared $K_n$--band photometry of a sample of $33$
edge--on galaxies, $30$ of which were studied kinematically by
\citeauthor{bf99}. The objects are
characterised by their bulge morphology, with $24$ galaxies showing a
b/p structure and the remaining a more spheroidal shape.
%
%

We applied the technique of unsharp masking to our images to analyse
their morphology. The results were then compared to edge--on
projections of b/p structures obtained from orbital calculations and
presented in Patsis, Skokos, and Athanassoula\ (\citeyear{psa02},
hereafter \citeauthor{psa02}). In addition, to quantify the b/p
structure in an objective manner, we studied our sample with the
fitting method described by \citeauthor{am02}, where generalised
gaussians of the form $I=I_0\exp[-(z/z_0)^\lambda]$ are used to fit
the vertical surface brightness profiles of the disk/bar components of
the simulations. To minimise the influence of the bulge in our images
(as the bulge was excluded from the N--body profiles), we apply a
magnitude cut to the fit in the inner parts of the galaxies.
\section{Results} 
\begin{figure}
\tabcapfont
\centerline{%
\begin{tabular}{c@{\hspace{0pc}}c}
\includegraphics[width=10.5pc,angle=0]{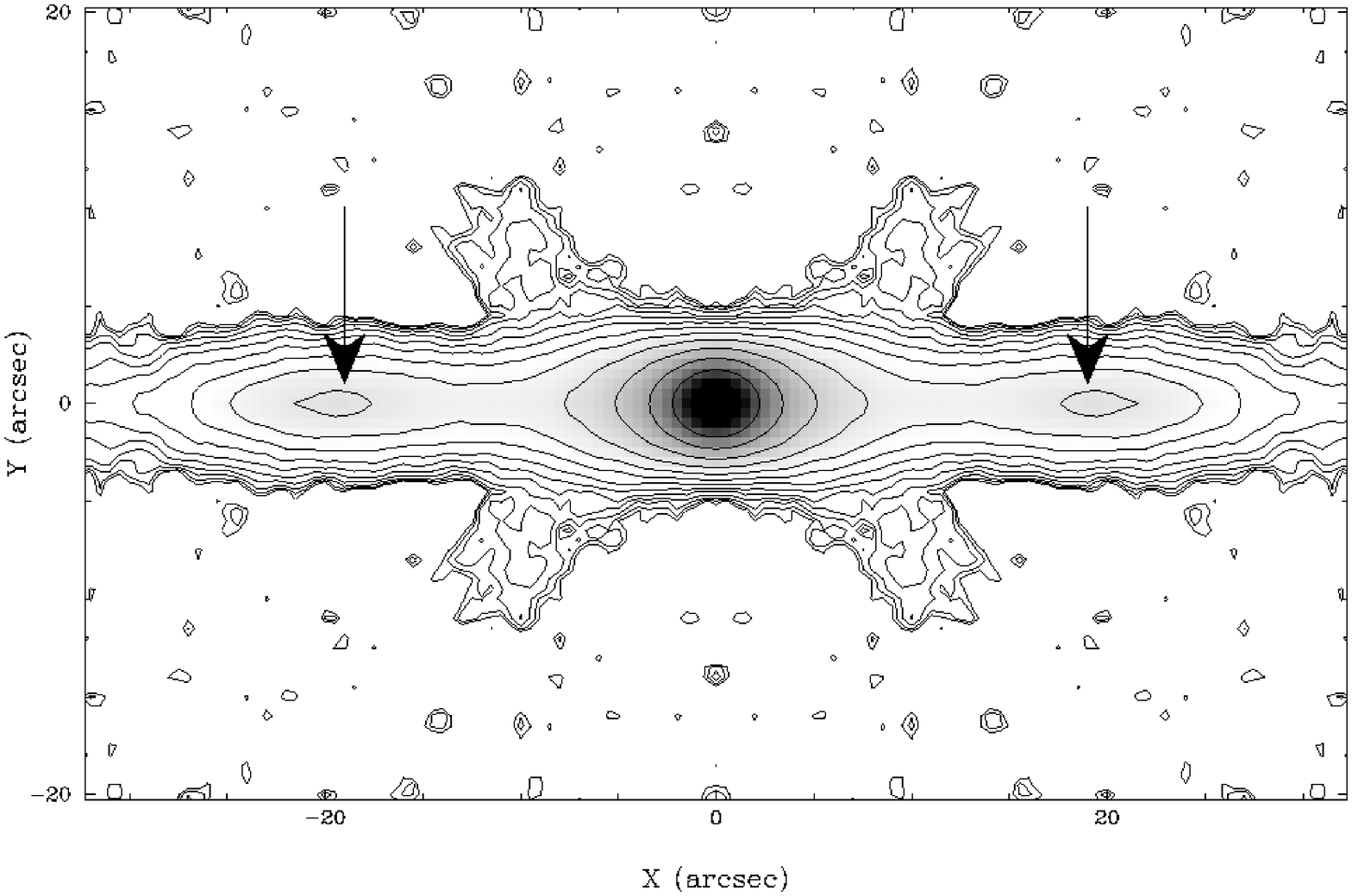} &
\includegraphics[width=10.5pc,angle=0]{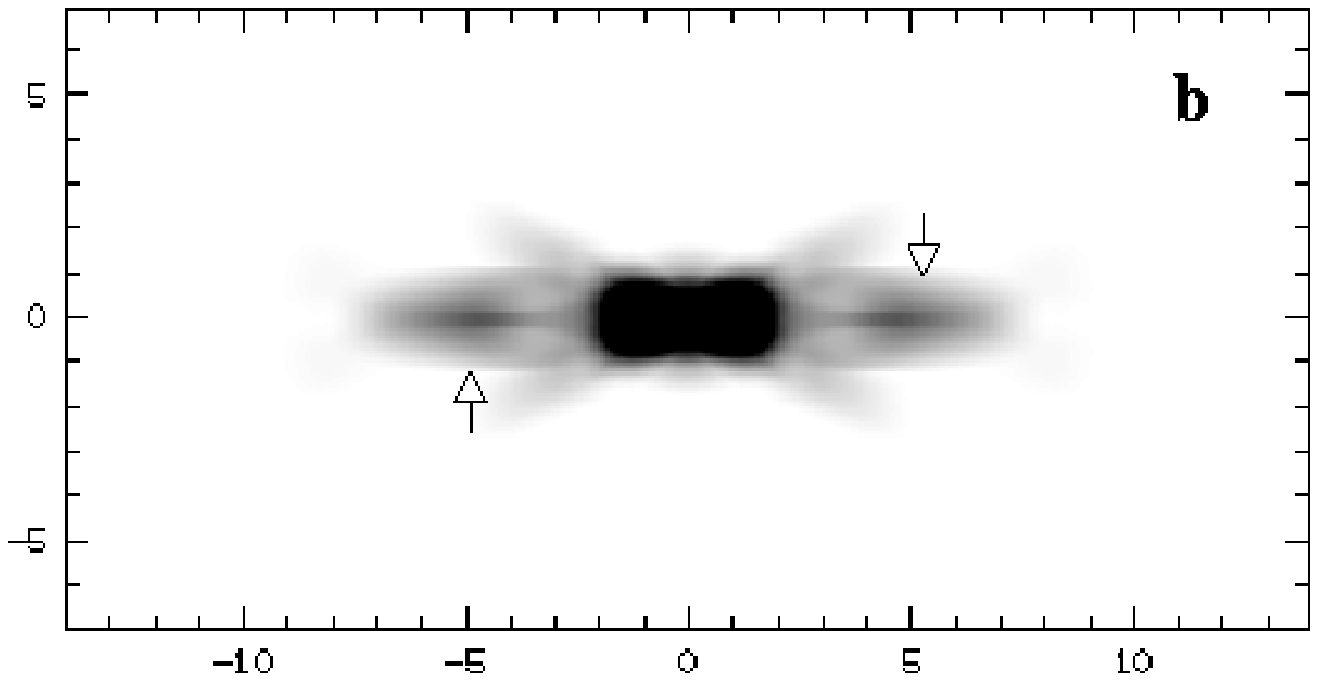} \\ 
a)~Unsharp masking of ESO~597~-036. & b)~Model A2 of PSA02.
\end{tabular}}
\caption[]{{\bf a)} Unsharp-masked image of ESO~597~-036. The galaxy
was symmetrised with respect to both symmetry axes to improve the
signal--to--noise ratio and a median filter of $16.5$~arcsec was
applied. The `branches' of the X--feature originate well outside of
the galactic center. Note again the local density enhancements in the
equatorial plane, at a projected radius of $20$~arcsec from the
centre. {\bf b)} Model A2 from \protect\citeauthor{psa02}. This side--on
profile is built from the superposition of the x2v1, x1v1, x1v3, and
x1v4 families of periodic orbits. The arrows indicate local density
enhancements. Reproduced from \protect\citeauthor{psa02} with the permission
of MNRAS published by Blackwell Science.}
\label{fig:enhance}
\end{figure}

Most of the original (e.g.\ Fig.~\ref{fig:contour}) and unsharp-masked
(e.g.\ Fig.~\ref{fig:enhance}a) images show local surface brightness
enhancements along the major axis on both sides of the central
regions. The positions of these characteristic enhancements coincide
with the ends of the bar (determined kinematically) and are observed
in $18$ of the $24$ galaxies showing a b/p `bulge', but in only $1$
of the remaining $9$ non--b/p galaxies. A possible explanation is an
inner ring delineating the ends of the bar, since the edge--on
projection would enhance the surface brightness at those positions
(see \citeauthor{am02}). A similar effect can also occur in the case
of spiral arms starting at the bar's ends, but an alternative
explanation has been proposed by \citeauthor{psa02}, who explain the
enhancements by a combination of four different orbit families and the
projection of material trapped around stable periodic orbits close to
the bar's ends (Fig.~\ref{fig:enhance}b). 
%
%

In addition to the surface brightness maxima, the apparent X--shape of
b/p `bulges' is clearly seen in the unsharp--masked images of the sample
galaxies (see, again, Fig.~\ref{fig:enhance}a).  Two types of
X--shapes are present in our sample. The first type is characterised
by the fact that the `branches' of the X--feature originate at the
center of the galaxy. In the other cases, the branches seem to
originate well outside of the central regions (e.g.\
Fig.~\ref{fig:enhance}a). \citeauthor{psa02} also find these two
separate morphologies in their work. For example, their model A2, with
a relatively slow pattern speed, reproduces an X--morphology with
branches emerging from the centre of the galaxy
(Fig.~\ref{fig:enhance}b), whereas their model D, characterised by a
very strong bar, has an X--feature with branches starting away from
the central bulge (as in Fig.~\ref{fig:enhance}.a). A comparison with
unsharp--masked images is justified here because \citeauthor{psa02}'s
models include only periodic orbits, which are representative building
blocks of the galaxies but not a self-consistent model.

\begin{figure}[H] 
\tabcapfont
\centerline{%
\begin{tabular}{c@{\hspace{1pc}}c}
\includegraphics[width=12.0pc,angle=0]{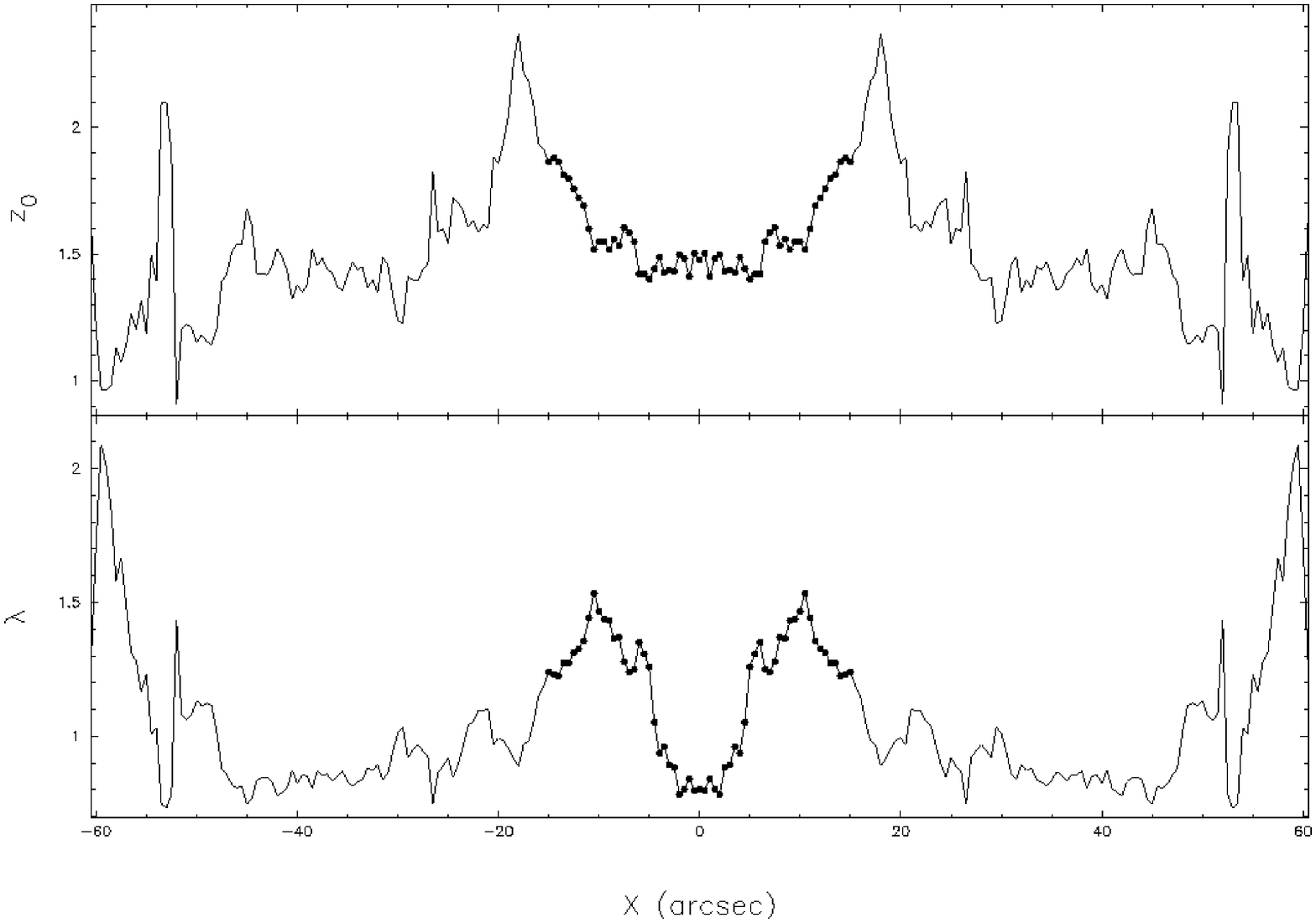} &
\includegraphics[width=15.0pc,angle=0]{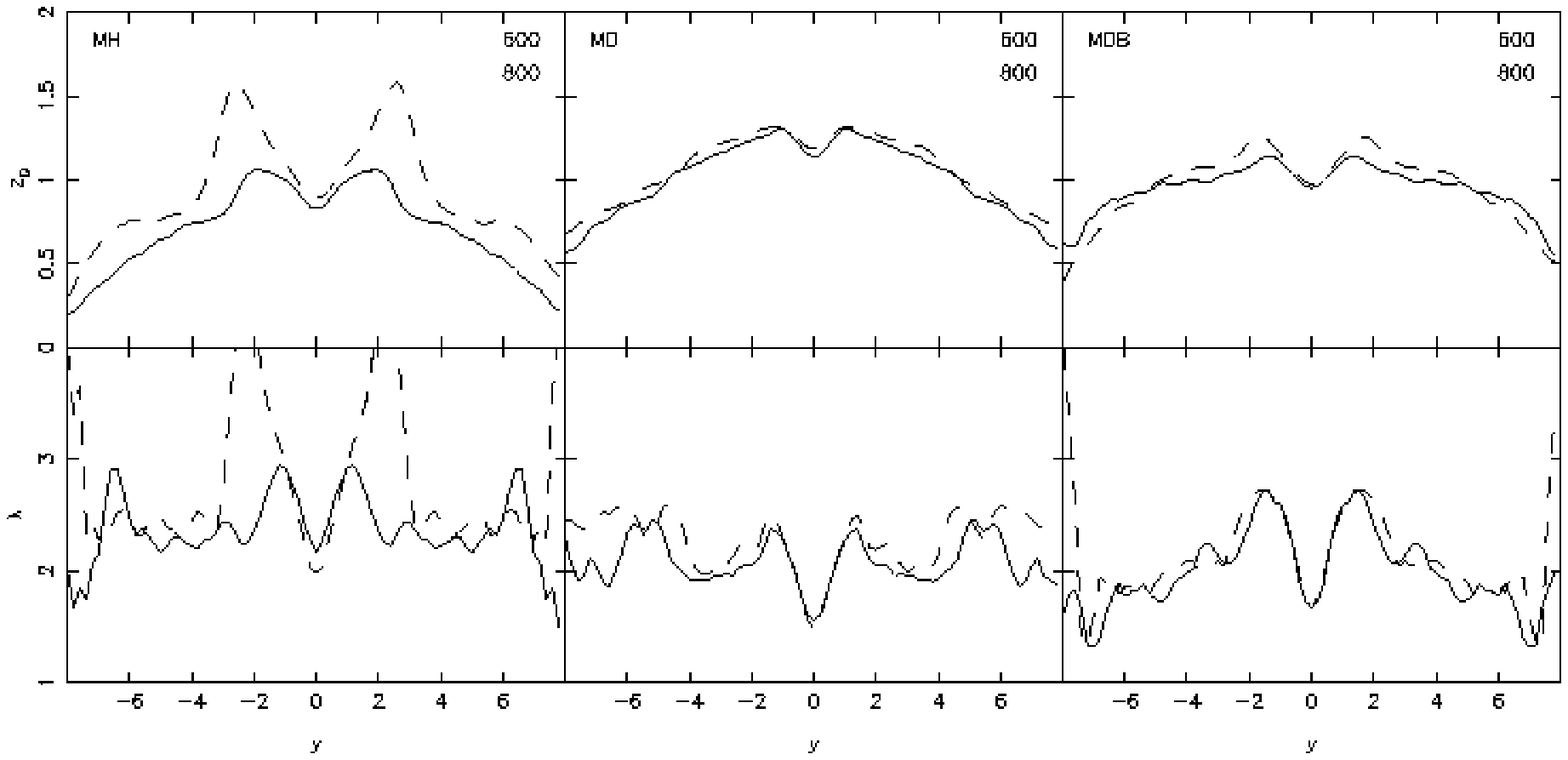} \\ 
a)~Vertical fits to ESO~443~-042. & b)~Vertical fits to AM02's models.
\end{tabular}}
\caption[]{{ \bf a)} Vertical fits to ESO~443~-042. Generalised
gaussians were fit to vertical cuts of a symmetrised image. The inner
parts are excluded from the fit to minimise the bulge contribution and
values obtained at those positions are marked with dots. Note that
despite the exclusion of the central parts the bulge still contributes
somewhat within $15$~arcsec. The upper panel shows the scaleheight
$z_0$ and the lower one the shape parameter $\lambda$, both as a
function of the projected radius. {\bf b)} Same as a) but for the
N--body simulations discussed by \protect\citeauthor{am02}. Each column
corresponds to a different model. From left to right: model MH
(massive halo), MD (massive disk), and MDB (massive disk with bulge).
Note, however, that in the last case the bulge was not taken into
account in the fits. Solid and dashed lines correspond to times $600$
and $800$, respectively. Reproduced from \protect\citeauthor{am02} with the
permission of MNRAS published by Blackwell Science.}
\label{fig:vertfig} \end{figure}

The quantification of the peanut structure of ESO~443~-042 by vertical
fits (Fig.~\ref{fig:vertfig}a) shows a qualitative similarity with the
fits of model MH in \citeauthor{am02} (massive halo, time $800$;
Fig.~\ref{fig:vertfig}b). In both cases the scaleheight $z_0$ and the
shape parameter $\lambda$ show a clear peak at or near the maximal
extension of the b/p structure. But the two peaks are in fact at
slightly different radii; in both the observations and the N--body
simulations, we can observe that the peaks of the shape parameter
$\lambda$ are closer to the center than those of the scaleheight
$z_0$. The analysis of the fits for the entire sample is ongoing.
%
%
\vspace{-0.4cm}
%
\acknowledgements {\footnotesize GA thanks the Laboratoire
d'Astrophysique de Marseille for support during visits when essential
parts of this work were done and acknowledges financial support from
the Deutscher Akademischer Austausch Dienst (grant D/02/02273). EA and
AB would like to thank the IGRAP, the Region PACA, the INSU/CNRS, and
the University of Aix--Marseille~I for funds to develop the GRAPE and
BEOWULF computing facilities used for the simulations. Support for
this work was also provided by NASA through Hubble Fellowship grant
HST-HF-01136.01 awarded by the Space Telescope Science Institute,
which is operated by the Association of Universities for Research in
Astronomy, Inc., for NASA, under contract NAS~5-26555.}
\vspace{-0.2cm}
%

%
\end{article}
\end{document}